\documentclass[english,aps,prl,twocolumn,sort&compress]{revtex4}
\usepackage[T1]{fontenc}
\usepackage[latin9]{inputenc}
\usepackage{amsthm}
\usepackage{amsmath}
\usepackage{graphicx}
\usepackage{esint}

\makeatletter
\@ifundefined{textcolor}{}
{%
 \definecolor{BLACK}{gray}{0}
 \definecolor{WHITE}{gray}{1}
 \definecolor{RED}{rgb}{1,0,0}
 \definecolor{GREEN}{rgb}{0,1,0}
 \definecolor{BLUE}{rgb}{0,0,1}
 \definecolor{CYAN}{cmyk}{1,0,0,0}
 \definecolor{MAGENTA}{cmyk}{0,1,0,0}
 \definecolor{YELLOW}{cmyk}{0,0,1,0}
 }

\makeatother

\usepackage{babel}

\begin{document}

\title{Active Biopolymers Confer Fast Reorganization Kinetics}

\author{Douglas Swanson}

\email{dsswanso@princeton.edu}

\affiliation{Department of Physics, Princeton University, Princeton, New Jersey,
USA}

\author{Ned S. Wingreen}

\email{wingreen@princeton.edu}

\affiliation{Department of Molecular Biology, Princeton University, Princeton,
New Jersey, USA}
\begin{abstract}
Many cytoskeletal biopolymers are \textquotedblleft{}active,\textquotedblright{}
consuming energy in large quantities. In this Letter, we identify
a fundamental difference between active polymers and passive, equilibrium
polymers: for equal mean lengths, active polymers can reorganize faster
than equilibrium polymers. We show that equilibrium polymers are intrinsically
limited to linear scaling between mean lifetime and mean length, $\mathrm{MFPT}\sim\left\langle L\right\rangle $,
by analogy to 1-d Potts models. By contrast, we present a simple active-polymer
model that improves upon this scaling, such that $\mathrm{MFPT}\sim\left\langle L\right\rangle ^{1/2}$.
Since to be biologically useful, structural biopolymers must typically
be many monomers long, yet respond dynamically to the needs of the
cell, the difference in reorganization kinetics may help to justify
active polymers' greater energy cost. PACS numbers: 87.10.Ed, 87.16.ad,
87.16.Ln
\end{abstract}
\maketitle
Cytoskeletal polymers play a key role in cellular reproduction, locomotion,
and transport \cite{Desai 1997,Pollard 2009,Cabeen 2010}. Biopolymers
like actin filaments and microtubules in eukaryotes and FtsZ, MreB,
and ParM in prokaryotes grow by accumulating monomers bound to the
nucleotide triphosphates ATP or GTP. The monomers hydrolyze these
triphosphates to the diphosphates ADP or GDP, consuming energy in
an irreversible process and inducing conformational changes that destabilize
the polymers. In some cells, cytoskeletal ATP consumption can approach
50\% of total cellular ATP consumption \cite{Bernstein 2003,Daniel 1986}.
What advantage do active polymers offer over passive, equilibrium
polymers to justify this costly energy expenditure?

We highlight a fundamental difference between active and equilibrium
polymers: active polymers can reorganize faster than equilibrium polymers.
Moreover, this difference in reorganization times widens as mean polymer
length grows. Since biological structures like mitotic spindles or
pseudopods must reach a certain size to accomplish their function,
yet be quickly deconstructed and reorganized, this intrinsic difference
may at least partly justify active polymers' greater energy cost.

A large class of equilibrium models describes a polymer as an ordered
sequence of monomers, each of one of $q$ types (including different
conformational states of the same molecule). Monomers can attach,
detach, and potentially interconvert among the $q$ types. Interactions
between neighboring monomers $\left\{ i,i+1\right\} $ contribute
free energy $J_{\left\{ i,i+1\right\} }$ to the total free energy
of the polymer. Such models can also describe polymers consisting
of bundles of $k$ protofilaments, by increasing the number of \textquotedblleft{}monomer\textquotedblright{}
types to $q^{k}$. These models are generalizations of 1-d, $q^{k}$-state
Potts models \cite{Wu 1982}. The free energy of an equilibrium polymer
in these models scales as the polymer length $L$ for large $L$,
specifically $\mathcal{F}\approx L\lambda_{\mathrm{max}}$, where
$\lambda_{\mathrm{max}}$ is the largest eigenvalue of the transfer
matrix \cite{Chaikin-Lubensky}. Hence, the equilibrium distribution
of polymer lengths will be exponential, $p(L)\sim e^{-L/\left\langle L\right\rangle }$,
with a characteristic mean length $\left\langle L\right\rangle =\frac{kT}{\lambda_{\mathrm{max}}}$.
Because for large $L$ the free energy effectively depends on only
the largest eigenvalue $\lambda_{\mathrm{max}}$, the dynamics are
essentially one-dimensional even for polymer bundles. This means that
the effective force $-\frac{d\mathcal{F}}{dL}\approx-\lambda_{\mathrm{max}}$
is constant, generating a constant negative-velocity drift in the
polymer length, with drift velocity $v_{d}\propto-\lambda_{\mathrm{max}}$.
(Polymers maintain a finite equilibrium distribution because this
negative drift is balanced by diffusion and nucleation of new polymers.)
Importantly, since the polymer length drifts towards zero at constant
negative drift velocity, starting from the nucleation length $L_{\mathrm{nucl}}$,
the mean polymer lifetime scales as $\frac{L_{\mathrm{nucl}}}{\left|v_{d}\right|}\propto\frac{1}{\lambda_{\mathrm{max}}}\propto\left\langle L\right\rangle $.
Thus the mean polymer lifetime or \textquotedblleft{}mean first-passage
time\textquotedblright{} (MFPT) scales as the mean length $\left\langle L\right\rangle $
for equilibrium polymers. This linear scaling is a fundamental limit
for an equilibrium polymer. In order to improve upon it, a biological
system must employ active or out-of-equilibrium processes. As an example,
we present a simple active-polymer model based on microtubule dynamics
that yields $\mathrm{MFPT}\sim\left\langle L\right\rangle ^{1/2}$.

\begin{figure}
\includegraphics[scale=0.3]{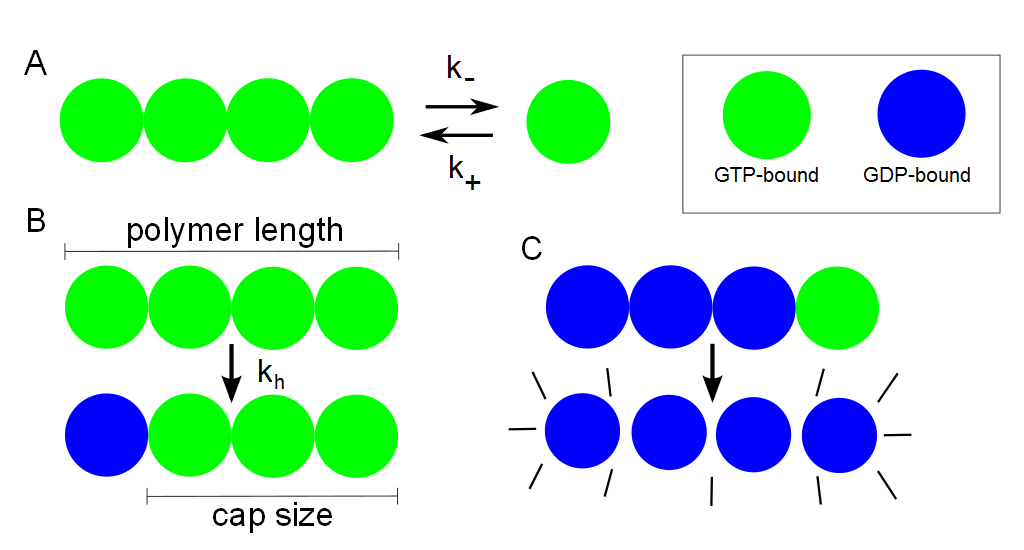}\caption{Schematic of growth and disassembly of a model active polymer. (A)
GTP-bound monomers (green) bind and unbind at the front end of the
polymer. (B) The last GTP-bound monomer at the back of the {}``cap''
undergoes hydrolysis to become an GDP-bound monomer (blue), decreasing
the cap size by one. (C) If the cap size shrinks to zero, the polymer
completely disassembles.\label{fig:Schematic-of-growth}}

\end{figure}

Microtubule growth and disassembly dynamics have been well-studied
\cite{Mitchison 1984,Bayley 1989,Leibler 1992}. In microtubules (and
ParM), GTP hydrolysis leads to stochastic rapid disassembly of the
entire polymer in a process called dynamic instability; the classic
experimental results \cite{Walker 1988} are reviewed in \cite{Desai 1997}.
Recent detailed models aim to explain specific aspects of the experimental
data \cite{Nedelec 2009,Antal 2007,Hinow 2009,Janosi 2002,Ranjith 2009}.
We consider instead a minimal microtubule model \cite{Leibler 1994}
that incorporates dynamic instability. Specifically, we model an active
polymer as an ordered sequence of monomers, each of which is bound
either to GTP or GDP (Fig. \ref{fig:Schematic-of-growth}). We call
the group of GTP-bound monomers at the front of the polymer the \textquotedblleft{}cap\textquotedblright{}
and denote its size by $x$. We denote the total number of monomers
(the polymer length) by $L$. GTP-bound monomers bind and unbind at
the front end of the polymer with rates $k_{+}$ and $k_{-}$, respectively
(Fig. 1A). GTP-bound monomers at the back of the polymer cap undergo
hydrolysis to become GDP-bound monomers with rate $k_{h}$ (Fig. 1B).
If the cap size shrinks to zero, the polymer completely disassembles
(Fig. 1C). New polymers of length and cap size 2 are nucleated with
rate $k_{\mathrm{nucl}}$. We call the concentration of free GTP-bound
monomers $c$ and analytically treat the mean-field regime where $c$
is constant, a good approximation for eukaryotic cells where the number
of monomers is typically large ($\sim10^{6}$). For comparison, we
also consider an equilibrium polymer that obeys the same rules but
without hydrolysis so that its length and cap size are equal. We show
explicitly that this particular equilibrium model satisfies the general
equilibrium scaling relation $\mathrm{MFPT}\sim\left\langle L\right\rangle $.

The exact master equation for the concentration $C_{L,x}$ of polymers
of length $L$ and cap size $x$ is\begin{eqnarray}
\frac{d}{dt}C_{L,x}= & k_{+}c\left(C_{L-1,x-1}-C_{L,x}\right)+k_{-}\left(C_{L+1,x+1}-C_{L,x}\right)\nonumber \\
 & +k_{h}\left(C_{L,x+1}-C_{L,x}\right)+k_{\mathrm{nucl}}c^{2}\delta_{L,2}\delta_{x,2}\,.\label{eq:master-eq}\end{eqnarray}
Coarse-graining this equation leads to a continuum Fokker-Planck (FP)
description of the probability $p=p(x,L,t)$ that a single active
polymer will have length $L$ and cap size $x$ at a time $t$ after
its birth: \begin{equation}
\frac{\partial p}{\partial t}=D\frac{\partial^{2}p}{\partial x^{2}}+D_{LL}\frac{\partial^{2}p}{\partial L^{2}}+D_{xL}\frac{\partial^{2}p}{\partial x\,\partial L}-a\frac{\partial p}{\partial x}-g\frac{\partial p}{\partial L}\,,\label{eq:fokker-planck}\end{equation}
where $D=D_{xx}=\frac{1}{2}\left(k_{+}c+k_{-}+k_{h}\right)$, $D_{LL}=\frac{1}{2}\left(k_{+}c+k_{-}\right)$,
and $D_{xL}=k_{+}c+k_{-}$ are diffusion coefficients, $a=k_{+}c-k_{-}-k_{h}$
is the cap drift velocity ($a<0$), and $g=k_{+}c-k_{-}$ is the length
drift velocity. The FP equation (\ref{eq:fokker-planck}) describes
the time-evolution of an individual polymer born at time 0 (the time
coordinate $t$ represents polymer age). In the mean-field regime
each polymer evolves independently once nucleated, and hence the $k_{\mathrm{nucl}}$
term does not appear (for details see \cite{Supp. Mat.}). Assuming
$g\gg-a$, the effect of cap diffusion dominates the effect of length
diffusion, and so we may neglect the $D_{LL}$ and $D_{xL}$ terms
\cite{Supp. Mat.}. Solving the FP equation (\ref{eq:fokker-planck})
yields\begin{equation}
p(x,L,t)=\frac{1}{\sqrt{4\pi Dt}}e^{\frac{-(x-at-2)^{2}}{4Dt}}\left(1-e^{\frac{-2x}{Dt}}\right)\delta(L-gt-2)\,,\label{eq:p(x,t)}\end{equation}
where the initial condition is $p(x,L,0)=\delta(x-2)\delta(L-2)$,
\textit{i.e.} polymers nucleate with length and cap size 2, and the
boundary condition is $p(0,L,t)=0$, \textit{i.e.} polymers with cap
size zero disassemble. (Changing the nucleation size does not affect
any essential results.)

The distribution of polymer lifetimes or \textquotedblleft{}first-passage
times\textquotedblright{} (FPTs) is\begin{equation}
P_{\mathrm{FPT}}(t)=-\frac{d}{dt}\iint_{0}^{\infty}p(x,L,t)\, dx\, dL=\frac{1}{\sqrt{\pi Dt^{3}}}e^{-\frac{\left(at+2\right)^{2}}{4Dt}}\,,\label{eq:FPT(t)}\end{equation}
and the mean first-passage time (MFPT) has the simple form $\mathrm{MFPT}=-2/a$.
The steady-state polymer length distribution is \begin{equation}
P_{\mathrm{active}}(L)=\iint_{0}^{\infty}p(x,L,t)\, dx\, dt\,,\label{eq:N_active(L)}\end{equation}
yielding the active-polymer average length \begin{equation}
\left\langle L\right\rangle _{\mathrm{active}}=\int_{0}^{\infty}P_{\mathrm{active}}(L)\, L\, dL=\frac{g(D-a)}{a^{2}}\,,\label{eq:active-<L>}\end{equation}
so that for long polymers $\left(\left\langle L\right\rangle \gg1,\left|a\right|\ll1\right)$,
one finds $\left\langle L\right\rangle \simeq gD/a^{2}$, and therefore
$\mathrm{MFPT}=-2/a\sim\left\langle L\right\rangle ^{1/2}$ to leading
order. This sublinear scaling requires some non-equilibrium process,
here the irreversible hydrolysis of monomers. Note that the details
of the non-equilibrium model do matter to the degree of sublinearity;
for example, a more realistic microtubule model \cite{Leibler 1994}
yields $\mathrm{MFPT\sim\left\langle L\right\rangle ^{0.7}}$ \cite{Supp. Mat.}.

By comparison, in the equilibrium limit of this model the cap is the
entire polymer. Dropping the $d/dL$ terms in (\ref{eq:fokker-planck})
yields an equilibrium solution that looks like (\ref{eq:p(x,t)})
without the factor $\delta(L-gt-2)$. The equilibrium length distribution
can then be obtained by integrating over polymer age:\begin{equation}
P_{\mathrm{equil}}(L)=-\frac{a}{D}e^{aL/D}\,,\label{eq:P_equilibrium(L)}\end{equation}
and the equilibrium-polymer average length is $\left\langle L\right\rangle _{\mathrm{equil}}=-D/a$.
Hence, we recover the linear scaling $\mathrm{MFPT}\sim\left\langle L\right\rangle $
expected for equilibrium polymers.

To validate these scaling relations, the master equation (\ref{eq:master-eq})
was simulated using the Gillespie algorithm \cite{Gillespie 1977}.
The monomer addition rate $k_{+}$ was held fixed throughout the simulations.
For equilibrium polymers we set $k_{h}=0$, while for active polymers
for simplicity we set $k_{-}=0$. The nucleation rate $k_{\mathrm{nucl}}$
was tuned to hold the steady-state fraction of polymerized material
constant at 75\%, with $k_{\mathrm{nucl}}$ from $10^{-5}$ to $10^{-8}$.
This is in line with experimentally measured values for microtubules
\cite{Beertsen 1982}. (Changing the fraction of polymerized material
to 95\% does not affect the qualitative results \cite{Supp. Mat.}.)
This leaves a single free parameter, $k_{-}$ for equilibrium polymers
and $k_{h}$ for active polymers, to control the polymer length and
MFPT. With only a single free parameter, each system is fully constrained
by holding either the MFPT or the average length fixed, thus yielding
a fair comparison between the equilibrium and active polymers.

Figure 2A shows the MFPT of the equilibrium and active polymers as
functions of their average length. The data points are from simulations
using 400,000 monomers; the curves are from $\mathrm{MFPT}=-2/a$
combined with $\left\langle L\right\rangle _{\mathrm{equil}}=-D/a$
and (\ref{eq:active-<L>}) for $\left\langle L\right\rangle _{\mathrm{active}}$.
The MFPT scales $\sim\left\langle L\right\rangle $ for equilibrium
polymers and $\sim\left\langle L\right\rangle ^{1/2}$ for active
polymers as expected. Hence, for the same average length, the active
polymers have much shorter mean lifetimes than the equilibrium polymers,
and this difference widens as average length grows. Figure 2B compares
length distributions for the equilibrium and active polymers with
the same MFPT ($\simeq$10). Theoretical results from (\ref{eq:N_active(L)})
and (\ref{eq:P_equilibrium(L)}) are shown in black. Agreement between
simulation and theory is excellent, validating our use of the FP equation.
(For a comparison of length distributions with the same $\left\langle L\right\rangle $,
see \cite{Supp. Mat.}.)

What might be the biological consequences of the different equilibrium
and active polymer scaling relations? To address this question, we
examine the time scale for large-scale spatial reorganization of structures,
\textit{i.e.} the time needed for a system to disassemble polymers
at one site, move the material to another site, and reassemble new
polymers. Cells often accomplish large-scale polymer reorganization
\textit{in vivo} by spatially regulating nucleation \cite{Machesky 1999,Oakley 1989,Malikov 2004}.
To model such regulation simply, we consider two spatial sites. We
start simulations with nucleation occurring only at site 1, allow
the system to come to steady state, then switch off nucleation at
site 1 and switch on nucleation at site 2. Monomers are assumed to
transition between the two sites with a \textquotedblleft{}diffusion\textquotedblright{}
rate $k_{D}$, while polymers do not diffuse. We define the {}``reorganization
time'' as the time needed after the nucleation switch for 50\% of
the final steady-state amount of polymerized material to assemble
at site 2. Initially we assume diffusion to be fast ($k_{D}=\infty$)
so that a single effective pool of free monomers is shared between
the two sites, and then we consider the more biologically-relevant
finite-diffusion regime.

\begin{figure}
\includegraphics[scale=0.32]{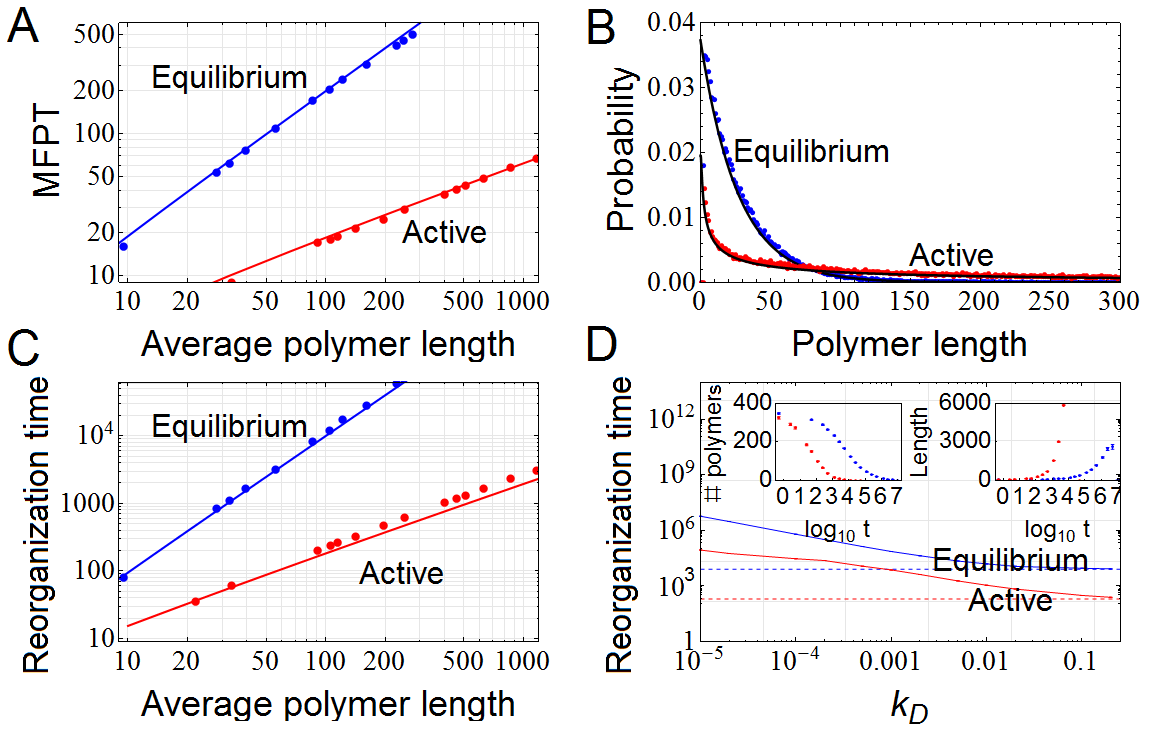}\caption{Lifetimes versus length for equilibrium (blue) and active (red) polymers.
(A) Mean first-passage time (MFPT) to polymer disassembly versus average
length. The monomer addition rate $k_{+}$ is held fixed, and time
is always measured in units of $(k_{+}c)^{-1}$, the mean time for
monomer binding. Solid lines are fits to theory with slopes 1 and
0.5 for equilibrium and active polymers, respectively. Some error
bars are smaller than data-point symbols. (B) Length distributions
with MFPT$\simeq$10, with theoretical fits (black) from (\ref{eq:N_active(L)})
and (\ref{eq:P_equilibrium(L)}). The equilibrium polymers have average
length 30, while the active polymers have average length 600. (C)
Reorganization time (defined in text) versus average length. Solid
lines are theoretical fits from (\ref{eq:t-reorganization}) with
slopes 2 and 1 for equilibrium and active polymers, respectively.
(D) Reorganization time versus monomer diffusion constant. Dashed
lines show infinite-diffusion limits. Insets: time series showing
average polymer number (left) and average polymer length (right) at
site 1 versus time after nucleation is switched off. \label{fig:Lifetimes-versus-length}}

\end{figure}

Figure 2C shows the reorganization time for our simple equilibrium
and active polymers as functions of their average length, with fast
diffusion. The reorganization time scales $\sim\left\langle L\right\rangle ^{2}$
for the equilibrium polymers and $\sim\left\langle L\right\rangle $
for the active polymers. Hence, for a given average length, the active
polymers reorganize faster than the equilibrium polymers, and like
for the MFPT the difference widens with increasing average length.
The scaling exponents differ from those for the MFPT because the reorganization
time is dominated by a few very long-lived long polymers, whereas
the MFPT is dominated by many short-lived short polymers.

To understand the scaling relations in Fig. 2C, we estimate the reorganization
time analytically as half the {}``average material age,\textquotedblright{}
which we define as the average age of the polymers in a steady-state
snapshot of the system, weighted by the length of each polymer: \begin{equation}
\tau_{\mathrm{reorg}}\approx\frac{1}{2}\frac{\int_{0}^{\infty}\mathrm{age}(L)\, P(L)\, L\, dL}{\int_{0}^{\infty}P(L)\, L\, dL}\,.\label{eq:t-reorganization}\end{equation}
This average material age captures the amount of time an average monomer
spends in an active polymer before it turns over, and well-approximates
the reorganization time. Long polymers have more material than short
polymers, and so the average material age is weighted by the length
of each polymer. For our equilibrium polymers, $\mathrm{age}(L)=-L/a$,
which gives $\tau_{\mathrm{reorg}}^{\mathrm{equil}}\approx2D/a^{2}\sim\left\langle L\right\rangle ^{2}$.
Indeed, since the drift velocity $a\sim1/\left\langle L\right\rangle $
for any equilibrium polymer model as discussed above, $\mathrm{age}(L)\sim L^{2}$,
and thus the equilibrium reorganization time scales generally $\sim\left\langle L\right\rangle ^{2}$.
As for the MFPT, this scaling is a fundamental property of equilibrium
polymers. To improve upon it requires active energy-dissipation or
some non-equilibrium process. For example, our simple active polymers
have $\mathrm{age}(L)=L/g$ and hence $\tau_{\mathrm{reorg}}^{\mathrm{active}}\approx\frac{4\left(a^{2}-3aD+3D^{2}\right)}{3a^{2}(D-a)}$,
so that for large $\left\langle L\right\rangle $ with $\left|a\right|\ll1$,
$\tau_{\mathrm{reorg}}^{\mathrm{active}}\approx4D/a^{2}\sim\left\langle L\right\rangle $. 

Next we consider the effects of slow monomer diffusion. Figure 2D
shows the reorganization time versus the monomer transition rate $k_{D}$
between sites for equilibrium and active polymers with the same average
length ($\simeq$90) and fraction of polymerized material (75\%).
The fast-diffusion limits, which are reached for $k_{D}\simeq1$ (in
units of $k_{+}c$), are shown with dashed lines. Strikingly, the
active polymers still reorganize much faster than the equilibrium
polymers even when slow diffusion limits the reorganization time.
To understand this effect, consider the equilibrium polymer dynamics
after nucleation is switched off at site 1: polymers there disassemble
stochastically and, since diffusion is slow, the released monomers
typically rejoin other polymers at the same site. Hence the number
of polymers at site 1 drops rapidly while the average polymer length
grows (Fig. 2D, insets). Eventually site 1 has only a few very long
polymers. The equilibrium polymers remain in this state for a very
long time, exchanging monomers with the free monomer pool at site
1 while diffusion slowly drains the pool, until the polymers finally
disassemble. Therefore, the time needed for slow diffusion to move
half the monomers from site 1 to site 2 is a good rough approximation
for the equilibrium reorganization time \cite{Supp. Mat.}. In contrast,
active polymers do not release monomers to the free monomer pool except
by disassembly. After the switch in nucleation, the number of active
polymers at site 1 drops while the average length grows, similar to
the equilibrium polymers. However, the few long-lived active polymers
at site 1 then quickly accumulate and hydrolyze all the free monomers
there, and then all the polymers disassemble. The time to hydrolyze
all the monomers at site 1, plus the time for half of those monomers
to diffuse to site 2, is therefore a good rough approximation for
the active reorganization time \cite{Supp. Mat.}.

In summary, we find a fundamental difference between active and equilibrium
polymers: active polymers can reach a fixed mean length with faster
reorganization kinetics than equilibrium polymers. Very generally,
we show that equilibrium polymer lifetimes scale linearly with mean
length. In contrast, active polymer lifetimes can scale sublinearly,
for example as $\left\langle L\right\rangle ^{1/2}$ in a simple model
motivated by microtubules or as $\left\langle L\right\rangle ^{0.7}$
in a more realistic model \cite{Leibler 1994,Supp. Mat.}. Furthermore,
in our example the kinetic advantage of active polymers persists even
for slow monomer diffusion. In a dynamic cellular environment, this
kinetic advantage may help justify active polymers' greater energy
cost.

Our comparison of active and equilibrium polymers predicts one might
find equilibrium polymers in biological contexts where polymer turnover
is slow or structures rarely need to be reorganized. This may be the
case for eukaryotic intermediate filaments \cite{Yoon 2001} or for
the bacterial homolog crescentin \cite{Jacobs-Wagner 2009}. In addition,
the existence of proteins like formins and profilins that accelerate
actin polymerization suggests that kinetic regulation of active polymers
is important to cells. Finally, although the specific active model
we consider is most closely based on polymers like microtubules or
ParM that exhibit dynamic instability, our conclusions regarding accelerated
kinetics could also relate to actin networks for which branching plays
a role analogous to nucleation \cite{Pollard 2007}.

Our model neglects many complexities of real biopolymers; clearly
active polymers accomplish more than simply reaching a certain length
with a certain lifetime. We only suggest that fast reorganization
kinetics might be a general (and generally desirable) feature of the
active polymer systems that are ubiquitous in biology.

We thank William Bialek, Zemer Gitai, Joshua Shaevitz, and Sven van
Teeffelen for helpful suggestions. D.S. was supported by a National
Science Foundation Graduate Research Fellowship and N.S.W. by National
Science Foundation Grant No. PHY-0957573.

\end{document}


\title{Supplementary Information for:\\
Active Biopolymers Confer Fast Reorganization Kinetics}

\author{Douglas Swanson}

\email{dsswanso@princeton.edu}

\affiliation{Department of Physics, Princeton University, Princeton, New Jersey,
USA}

\author{Ned S. Wingreen}

\email{wingreen@princeton.edu}

\affiliation{Department of Molecular Biology, Princeton University, Princeton,
New Jersey, USA}

\maketitle
\setcounter{figure}{0} \renewcommand{\thefigure}{S\arabic{figure}}
\setcounter{equation}{0} \renewcommand{\theequation}{S\arabic{equation}}

Here we solve the full Fokker-Planck equation defined in the main
text, and justify the simplifications employed there. Next, we estimate
the reorganization time for polymers to disassemble at one spatial
site and reassemble at another spatial site, for slow monomer diffusion.
We then compare equilibrium and active-polymer length distributions
with the same average length, and also consider the result of varying
the steady-state fraction of polymerized material. Finally, we obtain
scaling relations for a more realistic model of microtubule dynamics
\cite{Leibler 1994}, and demonstrate that an equilibrium two-species
model satisfies our predicted equilibrium scaling relation.

\section{Fokker-Planck Derivation}

Here we solve the full Fokker-Planck equation (2) defined in the main
text, and justify the simplifications leading to (3). The Fokker-Planck
equation for the probability $p=p(x,L,t)$ of finding a polymer with
length $L$ and cap size $x$ at a time $t$ after the polymer's birth
is \begin{equation}
\frac{\partial p}{\partial t}=D_{xx}\frac{\partial^{2}p}{\partial x^{2}}+D_{LL}\frac{\partial^{2}p}{\partial L^{2}}+D_{xL}\frac{\partial^{2}p}{\partial x\,\partial L}-a\frac{\partial p}{\partial x}-g\frac{\partial p}{\partial L}\,,\label{eq:fokker-planck}\end{equation}
where $D_{xx}=\frac{1}{2}\left(k_{+}c+k_{-}+k_{h}\right)$, $D_{LL}=\frac{1}{2}\left(k_{+}c+k_{-}\right)$,
and $D_{xL}=k_{+}c+k_{-}$ are diffusion coefficients, $a=k_{+}c-k_{-}-k_{h}$
is the cap drift velocity ($a<0$), and $g=k_{+}c-k_{-}$ is the length
drift velocity. This equation does not contain a term proportional
to $k_{\mathrm{nucl}}$. The master equation (1) describes the time-evolution
of an entire system of polymers (the time coordinate $t$ represents
wall-clock time), whereas the Fokker-Planck equation (2) describes
the time-evolution of an individual polymer born at time 0 (the time
coordinate $t$ represents polymer age). In the mean-field regime
we treat, the growth of an individual polymer born at time 0 is completely
independent of the other polymers in the system (i.e. once nucleated,
each polymer evolves independently). Therefore, the nucleation rate
$k_{\mathrm{nucl}}$ does not explicitly affect the normalized distributions
of polymer lengths, caps, or lifetimes. However, the nucleation rate
does affect the steady-state amount of polymerized material and hence
the free monomer concentration $c$, in cases where $c$ is allowed
to vary.

\begin{figure}
\includegraphics[scale=0.4]{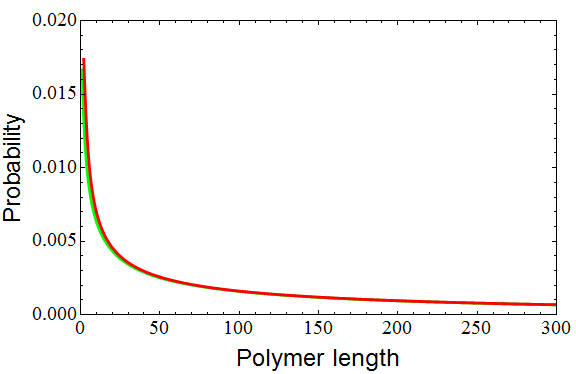}

\caption{\label{fig:s1}Comparison of active-polymer length distributions for
$\left\langle L\right\rangle \simeq132$. Length distributions were
calculated from Fokker-Planck theory using the full joint distribution
of polymer length and polymer age (\ref{eq:N_active(L)}) in red and
the delta-function approximation $p(L,t)=P_{\mathrm{alive}}(t)\cdot\delta(L-gt-2)$
used in the main text (5) in green.}

\end{figure}

The general solution to the Fokker-Planck equation (2) is \begin{eqnarray}
p(x,L,t)= & \frac{1}{2\pi t\sqrt{4D_{xx}D_{LL}-D_{xL}^{2}}}\exp\left[-\frac{D_{xx}(L-gt-2)^{2}+D_{LL}(x-at-2)^{2}-D_{xL}(L-gt-2)(x-at-2)}{\left(4D_{xx}D_{LL}-D_{xL}^{2}\right)t}\right]\left(1-e^{\frac{-2x}{D_{xx}t}}\right) & \,,\label{eq:p(x,L,t)}\\
\nonumber \end{eqnarray}
 which satisfies both the initial condition $p(x,L,0)=\delta(x-2)\delta(L-2)$
and the boundary condition $p(0,L,t)=0$. In the limit $D_{LL}=D_{xL}=0$,
this reduces exactly to (3). We note that we have solved (\ref{eq:fokker-planck})
subject to no boundary conditions on $L$. This is a valid assumption
for our active polymers, which nucleate with length $L_{\mathrm{nucl}}=2$
and then grow at average rate $g>0$, and therefore the effect of
the physical boundary condition at $L=0$ is negligible. More formally,
this assumption is valid provided $g\gg-a$, which implies the hydrolysis
rate $k_{h}\gg-a$. The probability that a polymer is still alive,
\textit{i.e.} has not disassembled, by age $t$ is \begin{equation}
P_{\mathrm{alive}}(t)=\int_{0}^{\infty}\int_{0}^{\infty}p(x,L,t)\, dx\, dL\,,\label{eq:P_alive}\end{equation}
 and the distribution of polymer lifetimes or {}``first-passage times''
is \begin{equation}
P_{\mathrm{FPT}}(t)=-\frac{d}{dt}P_{\mathrm{alive}}(t)=\frac{1}{\sqrt{\pi D_{xx}t^{3}}}e^{-\frac{(at+2)^{2}}{4D_{xx}t}}\,,\end{equation}
which is identical to (4) in the main text. Finally, in order to calculate
the steady-state polymer length distribution (5), we require the joint
distribution $p(L,t)$ between active polymer length $L$ and age
$t$. This may be obtained by integrating (\ref{eq:p(x,L,t)}) over
cap sizes: \begin{equation}
p(L,t)=\int_{0}^{\infty}p(x,L,t)\, dx\,,\end{equation}
 and then the length distribution is \begin{equation}
P_{\mathrm{active}}(L)=\int_{0}^{\infty}p(L,t)\, dt\,.\label{eq:N_active(L)}\end{equation}
 In the limit $D_{LL}=D_{xL}=0$ employed in the main text, the joint
distribution $p(L,t)=P_{\mathrm{alive}}(t)\cdot\delta(L-gt-2)$. Figure
\ref{fig:s1} shows a direct comparison of the length distributions
using both the full joint distribution (\ref{eq:N_active(L)}) in
red and the delta-function approximation $p(L,t)=P_{\mathrm{alive}}(t)\cdot\delta(L-gt-2)$
used in the main text (5) in green. The length distributions are nearly
identical, providing a practical validation for the intuitive simplifications
used to obtain (3)-(6) in the main text. Inserting the delta-function
approximation into (\ref{eq:N_active(L)}) gives \begin{equation}
P_{\mathrm{active}}(L)=\frac{1}{g}P_{\mathrm{alive}}\left(\frac{L}{g}\right)\end{equation}
 and the mean polymer length is calculated by integration by parts:
\begin{equation}
\left\langle L\right\rangle =\frac{\int_{0}^{\infty}L\, P_{\mathrm{active}}(L)\, dL}{\int_{0}^{\infty}P_{\mathrm{active}}(L)\, dL}=\frac{\int_{0}^{\infty}L\, P_{\mathrm{alive}}\left(\frac{L}{g}\right)\, dL}{\int_{0}^{\infty}P_{\mathrm{alive}}\left(\frac{L}{g}\right)\, dL}=\frac{1}{2}\frac{\int_{0}^{\infty}L^{2}\, P_{\mathrm{FPT}}\left(\frac{L}{g}\right)\, dL}{\int_{0}^{\infty}L\, P_{\mathrm{FPT}}\left(\frac{L}{g}\right)\, dL}=\frac{g(D-a)}{a^{2}}.\end{equation}

More formally, we may argue that the effect of cap diffusion dominates
the effect of length diffusion. Motivated by the scaling relations
found in the main text (and the fact that the average cap size $\left\langle x\right\rangle =-\frac{D}{a}\sim\left\langle L\right\rangle ^{1/2}$),
we make the ansatz that we may rescale $L$ to $\widetilde{L}=L/\left\langle L\right\rangle $
and $x$ to $\widetilde{x}=x/\left\langle L\right\rangle ^{1/2}$.
Then we obtain the non-dimensionalized Fokker-Planck equation: \begin{equation}
\frac{\partial p}{\partial t}=\frac{D_{xx}}{\left\langle L\right\rangle }\frac{\partial^{2}p}{\partial\widetilde{x}^{2}}+\frac{D_{LL}}{\left\langle L\right\rangle ^{2}}\frac{\partial^{2}p}{\partial\widetilde{L}^{2}}+\frac{D_{xL}}{\left\langle L\right\rangle ^{3/2}}\frac{\partial^{2}p}{\partial\widetilde{x}\,\partial\widetilde{L}}-\frac{a}{\left\langle L\right\rangle ^{1/2}}\frac{\partial p}{\partial\widetilde{x}}-\frac{g}{\left\langle L\right\rangle }\frac{\partial p}{\partial\widetilde{L}}\,.\end{equation}
 For long polymers ($\left\langle L\right\rangle \gg1$) the $D_{LL}$
and $D_{xL}$ terms will be much smaller than the remaining three
terms, and we may neglect the $D_{LL}$ and $D_{xL}$ terms. (More
generally, this approximation applies as long as $k_{h}$,$g\gg-a$.)
This leads immediately to the solution (3) in the main text.

\section{Polymer Reorganization Time}

Here we estimate the reorganization time for polymers to disassemble
at one spatial site (site 1) and reassemble at another spatial site
(site 2) following a switch in nucleation. As defined in the main
text, the reorganization time is the time needed after the switch
for half of the final steady-state amount of polymerized material
to assemble at site 2. For fast monomer diffusion, we estimated the
reorganization time in the main text as half the average material
age.

When diffusion is slow, we must estimate the reorganization time differently.
For equilibrium polymers, after the switch in nucleation, site 1 comes
to a local quasi-steady-state free monomer concentration: as diffusion
slowly drains the free monomer pool at site 1, depolymerization keeps
the free monomer concentration roughly constant. The quasi-steady-state
free monomer concentration $c_{1}$ is \begin{equation}
c_{1}=\frac{k_{-}}{k_{+}}\,.\end{equation}
At site 2, the free monomer concentration also quickly comes to a
value that stays roughly constant for a long time. However, this value
is smaller than $c_{1}$ since nucleation removes free monomers from
the pool at site 2. We may estimate this free monomer concentration
$c_{2}$ roughly by \begin{eqnarray}
k_{-}N_{2}-k_{+}c_{2}N_{2}-2k_{\mathrm{nucl}}c_{2}^{2} & = & 0\,,\\
k_{\mathrm{nucl}}c_{2}^{2}-\frac{N_{2}}{\mathrm{MFPT}} & = & 0\,,\end{eqnarray}
 where $N_{2}$ is the number of polymers at site 2. Although $N_{2}$,
like $c_{2}$, is time-dependent and slowly grows as diffusion moves
monomers from site 1 to site 2, immediately after the switch $N_{2}$
quickly comes to a value that stays roughly constant for a long time.
Using this value, which is approximately half the final expected steady-state
number of polymers, and using the final expected MFPT (which is the
same as before the switch in nucleation), we may come up with a rough
estimate for $c_{2}$. Then an estimate for the reorganization time
is just the time needed for half the final steady-state amount of
polymerized material to diffuse down the concentration \textquotedblleft{}gradient\textquotedblright{}
from site 1 to site 2. If $c_{\mathrm{steady}}$ is the final steady-state
concentration of polymerized material, then our estimate for the reorganization
time may be written as\begin{equation}
\tau_{\mathrm{reorg}}^{\mathrm{equil}}\approx\frac{c_{\mathrm{steady}}/2}{k_{D}(c_{1}-c_{2})}\,.\end{equation}
For typical parameter values (such as those used in Fig. 2D), we find
that this estimate for the equilibrium-polymer reorganization time
is accurate to within a factor of 2.

For active polymers, after the switch in nucleation, site 1 can replenish
the free monomer pool only by polymer disassembly. Since diffusion
is slow, monomers released by disassembly typically rejoin other polymers
at site 1. Hence after the switch, the number of active polymers at
site 1 rapidly drops while the average polymer length grows, until
eventually site 1 has only a few very long, very long-lived polymers.
These few long-lived polymers then quickly accumulate and hydrolyze
all the free monomers at site 1, and then all the polymers disassemble.
Therefore, a rough estimate for the reorganization time is just the
time needed to hydrolyze all the monomers at site 1, plus the time
needed for half of that material to diffuse to site 2:\begin{equation}
\tau_{\mathrm{reorg}}^{\mathrm{active}}\approx\frac{c_{\mathrm{total,1}}}{k_{h}}+\frac{1}{2k_{D}}\,,\end{equation}
where $c_{\mathrm{total},1}$ is the initial total concentration of
material at site 1. For typical parameter values (such as those used
in Fig. 2D), this estimate for the active-polymer reorganization time
is accurate to better than 10\%.

\section{Comparison of equilibrium and active-polymer length distributions
with the same average length}

\begin{figure}
\includegraphics[scale=0.4]{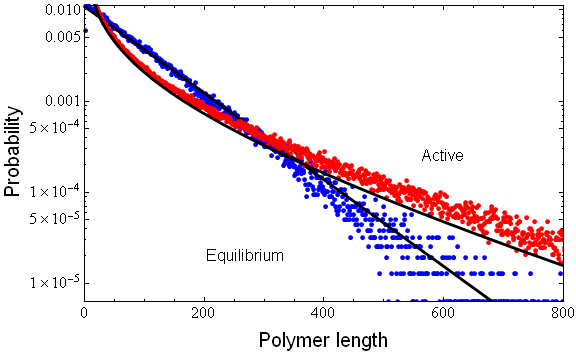}

\caption{\label{fig:s2}Histogram of lengths for equilibrium (blue) and active
(red) polymers with average length $\simeq90$. Black curves are theoretical
fits from (5) and (7).}

\end{figure}

In Fig. 2B, we showed that equilibrium and active polymers with the
same MFPT have qualitatively different length distributions: the active
polymers have a much longer average length. Figure \ref{fig:s2} shows
that even for equilibrium and active polymers with the same average
length, there are more very long active polymers than very long equilibrium
polymers.

\section{MFPT versus average length for 95\% polymerized material}

\begin{figure}
\includegraphics[scale=0.4]{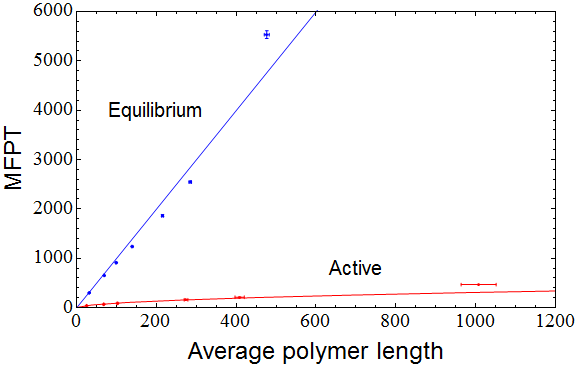}

\caption{\label{fig:s3}MFPT versus average length for equilibrium (blue) and
active (red) polymers, with the steady-state fraction of polymerized
material held at 95\%. Curves are theoretical fits using $\mathrm{MFPT}=\frac{2}{D}\left\langle L\right\rangle _{\mathrm{equil}}$
for equilibrium polymers and (6) for active polymers.}

\end{figure}

In the simulations described in the main text, we adjusted the nucleation
rate $k_{\mathrm{nucl}}$ so that the steady-state fraction of polymerized
material was always held fixed at 75\%. For comparison, Fig. \ref{fig:s3}
shows the MFPT of equilibrium and active polymers as functions of
their average length, akin to Fig. 2A, with the steady-state fraction
of polymerized material held instead at 95\%. The general trends are
the same as those seen in Fig. 2A: for a given average length, active
polymers disassemble significantly faster than equilibrium polymers
in a manner well-predicted by Fokker-Planck theory, and the difference
increases as average length increases. The MFPT is measured in the
same time units as defined in the main text, even though the steady-state
free monomer concentration is different.

\section{MFPT versus average length for partially-vectorial hydrolysis and
finite depolymerization velocity}

In the simulations described in the main text, we studied a minimal
microtubule model with vectorial hydrolysis (i.e. only the last monomer
in the polymer cap is subject to hydrolysis) and infinite depolymerization
velocity when the polymer cap size goes to zero. Here we relax these
assumptions. First, we consider the more realistic partially-vectorial
microtubule model described by Flyvbjerg, Holy, and Leibler (FHL)
\cite{Leibler 1994}. This model has a vectorial hydrolysis rate $k_{h}$
for any GTP-bound monomer adjacent to a GDP-bound monomer (i.e. identical
to our minimal model), and a smaller non-vectorial hydrolysis rate
$r=k_{r}/\delta x$ for all other GTP-bound monomers, representing
the rate per unit length at which new GTP-GDP interfaces are created.
$\delta x\simeq8\mathrm{nm}$ is the length scale of an individual
tubulin monomer. In all other respects the FHL model is identical
to our own.

\begin{figure}
\includegraphics[scale=0.5]{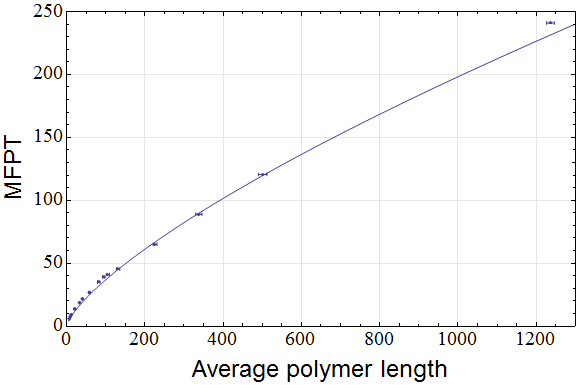}

\caption{\label{fig:s4}MFPT versus average length for the FHL model \cite{Leibler 1994}.
Curve is a numerical fit to a power law, $\mathrm{MFPT}\sim\left\langle L\right\rangle ^{0.7}$
.}

\end{figure}

Figure \ref{fig:s4} shows the MFPT versus average length for polymers
in this more realistic microtubule model, akin to Fig. 2A. In these
simulations $k_{r}=k_{h}/10$, roughly the value FHL found by fitting
to experimental data. The MFPT scales $\sim\left\langle L\right\rangle ^{0.7}$,
in between the $\sim\left\langle L\right\rangle ^{1/2}$ scaling of
our purely vectorial model and the $\sim\left\langle L\right\rangle $
equilibrium scaling.

\begin{figure}
\includegraphics[scale=0.5]{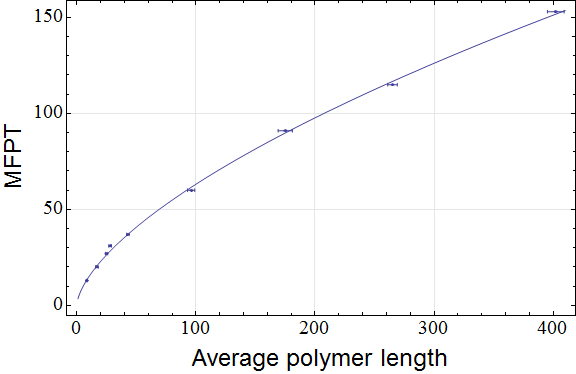}

\caption{\label{fig:s5}MFPT versus average length for the modified FHL model
\cite{Leibler 1994} with finite zero-cap depolymerization rate. Curve
is a numerical fit to a power law, $\mathrm{MFPT}\sim\left\langle L\right\rangle ^{0.7}$
.}

\end{figure}

Since the FHL paper did not consider microtubule disassembly dynamics
(the purpose of the paper was only to predict the distribution of
polymer catastrophe times), we next consider the effect of adding
to this model a finite depolymerization rate $k_{-}^{\mathrm{(dep)}}$
representing the rate of polymer disassembly after the cap size goes
to zero. We choose $k_{-}^{\mathrm{(dep)}}$ to be twice the polymerization
rate $k_{+}c$. All other parameters are the same as in the other
FHL simulations above. Figure \ref{fig:s5} shows the MFPT versus
average length in this modified FHL model. Again the MFPT scales $\sim\left\langle L\right\rangle ^{0.7}$.

Hence, the MFPT scales sublinearly with mean polymer length in a more
realistic non-equilibrium microtubule model. We did find two regimes
that produce linear scaling in the modified FHL model: if the average
depolymerization time for the entire polymer is much longer than the
average time for the cap size to reach zero, then the MFPT is dominated
by the depolymerization time and scales linearly with average length
(data not shown). However, this is not the regime observed experimentally
for microtubules \cite{Walker 1988}. Also, if the vectorial hydrolysis
rate goes to zero (i.e. the hydrolysis rate is the same for all GTP-bound
monomers), then the MFPT scales linearly with the average length (data
not shown), but this is not thought to be the case for microtubules.

\section{MFPT versus average length for two-species equilibrium polymer}

In the main text we argued that a large class of equilibrium polymer
models produces mean lifetimes that scale linearly with mean lengths.
This class includes all equilibrium polymers comprised of an ordered
sequence of monomers, each of one of $q$ types or conformations (or
$q^{k}$ types for $k$ protofilaments), where interactions between
neighboring monomers $\left\{ i,i+1\right\} $ contribute free energy
$J_{\left\{ i,i+1\right\} }$ to the total free energy of the polymer.
This class maps one-to-one onto 1-d, $q^{k}$-state Potts models,
and produces a free energy that scales linearly with mean length.

To provide further validation for our argument, we simulated one such
equilibrium model with $q=2$ different monomer types. This model
is designed to be the equivalent of our minimal non-equilibrium microtubule
model, but subject to the equilibrium constraint of reversibility.
Specifically, monomers of one type (the {}``GTP-bound'' type) can
attach and detach at the front of the polymer. Monomers at the back
of the cap can interconvert between the two types, and when the cap
size goes to zero (i.e. when the polymer is entirely composed of monomers
of the {}``GDP-bound'' type), monomers of the {}``GDP-bound''
type can both attach and detach.

\begin{figure}
\includegraphics[scale=0.5]{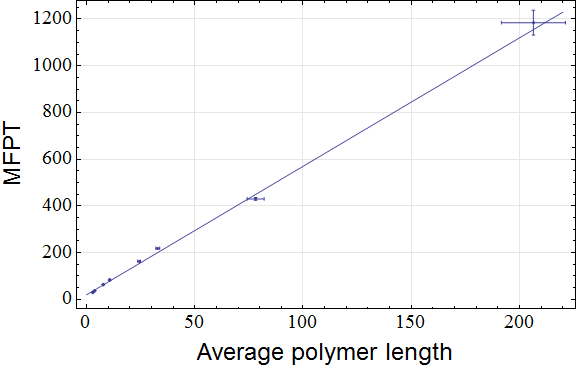}

\caption{\label{fig:s6}MFPT versus average length for the two-species equilibrium
model in the regime where one monomer type is energetically favored
over another. This regime is designed to mimic our non-equilibrium
model from the main text. Curve is a numerical fit to a linear relationship,
$\mathrm{MFPT}\sim\left\langle L\right\rangle $ .}

\end{figure}

First we consider an energetic regime comparable to our non-equilibrium
model. We set the energy of {}``GTP-bound'' monomers to favor attachment,
and that of {}``GDP-bound'' monomers to favor detachment. This means
that {}``GTP-bound'' monomers are more stable in polymers, and {}``GDP-bound''
monomers are more stable as free monomers. However, in order for monomer
interconversion to occur mainly from {}``GTP-bound'' to {}``GDP-bound'',
as for the non-equilibrium system, {}``GDP-bound'' monomers must
also be more stable than {}``GTP-bound'' monomers even in polymers.
This implies that the total monomer population will consist mostly
of {}``GDP-bound'' monomers. Indeed, when simulated in this regime,
with the same overall $k_{+}$ and fraction of polymerized material
as in our original simulations, we find a ratio of {}``GDP-bound''
monomers to {}``GTP-bound'' monomers of 10:1. Furthermore, in the
limit where monomer interconversion approaches unidirectional from
{}``GTP-bound'' to {}``GDP-bound'', the system approaches a single
species, and we recover our original equilibrium model. Finally, Fig.
\ref{fig:s6} shows the MFPT versus average length for this two-species
equilibrium model. The MFPT scales linearly with average length as
expected.

\begin{figure}
\includegraphics[scale=0.5]{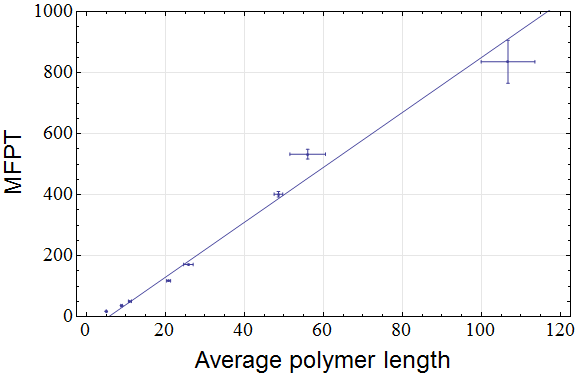}

\caption{\label{fig:s7}MFPT versus average length for the two-species equilibrium
model in the regime where the monomer types are energetically equivalent.
Curve is a numerical fit to a linear relationship, $\mathrm{MFPT}\sim\left\langle L\right\rangle $
.}

\end{figure}

Second we consider the regime where the two monomer species are energetically
equivalent. Figure \ref{fig:s7} shows the MFPT versus average length
for this model. Again the MFPT scales linearly with average length
as expected.

